\documentstyle[prl,epsfig,aps,floats,amstex]{revtex}  
\input psfig.sty
\def\ppbar{$p\overline{p}~$}             
\def\pt{$p_T$}                          
\def\met{\mbox{${\hbox{$E$\kern-0.6em\lower-.1ex\hbox{/}}}_T$}} 
\def\mex{\mbox{${\hbox{$E$\kern-0.6em\lower-.1ex\hbox{/}}}_x$}} 
\def\mey{\mbox{${\hbox{$E$\kern-0.6em\lower-.1ex\hbox{/}}}_y$}} 
\def\mexy{\mbox{${\hbox{$E$\kern-0.6em\lower-.1ex\hbox{/}}}_{x,y}$}} 
 
\def\invpb{${\rm pb^{-1}}$}
\def\intlumi{\int\!{\cal L}~dt}
\def\mll{M(l^+,l^-)}                     
\def\D0{D\O}                            
\font\eightit=cmti8
\def\r#1{\ignorespaces $^{#1}$}

\newcommand{\Wg}{$W\gamma$~}
\newcommand{\Zg}{$Z\gamma$~}

\begin{document}
\lefthyphenmin=2
\righthyphenmin=3

%
%
\title{
Measurement of {\boldmath $W\gamma$} and {\boldmath $Z\gamma$} Production in 
{\boldmath $p\overline{p}$} Collisions at\\ 
{\boldmath $\sqrt{s}$} = 1.96 TeV\\
}

\maketitle


\font\eightit=cmti8
\def\r#1{\ignorespaces $^{#1}$}
\hfilneg
\begin{sloppypar}
\noindent
D.~Acosta,\r {16} J.~Adelman,\r {12} T.~Affolder,\r 9 T.~Akimoto,\r {54}
M.G.~Albrow,\r {15} D.~Ambrose,\r {43} S.~Amerio,\r {42}  
D.~Amidei,\r {33} A.~Anastassov,\r {50} K.~Anikeev,\r {31}
% ...
\end{sloppypar}
\vskip .026in
\begin{center}
(CDF Collaboration)
\end{center}

\vskip .026in
\begin{center}
\r 1  {\eightit Institute of Physics, Academia Sinica, Taipei, Taiwan 11529, 
Republic of China} \\
\r 2  {\eightit Argonne National Laboratory, Argonne, Illinois 60439} \\
\r 3  {\eightit Institut de Fisica d'Altes Energies, Universitat Autonoma
de Barcelona, E-08193, Bellaterra (Barcelona), Spain} \\
\r 4  {\eightit Istituto Nazionale di Fisica Nucleare, University of Bologna,
I-40127 Bologna, Italy} \\
\r 5  {\eightit Brandeis University, Waltham, Massachusetts 02254} \\
\r 6  {\eightit University of California at Davis, Davis, California  95616} \\
\r 7  {\eightit University of California at Los Angeles, Los 
Angeles, California  90024} \\
\r 8  {\eightit University of California at San Diego, La Jolla, California  92093} \\ 
\r 9  {\eightit University of California at Santa Barbara, Santa Barbara, California 
93106} \\ 
\r {10} {\eightit Instituto de Fisica de Cantabria, CSIC-University of Cantabria, 
39005 Santander, Spain} \\
\r {11} {\eightit Carnegie Mellon University, Pittsburgh, PA  15213} \\
\r {12} {\eightit Enrico Fermi Institute, University of Chicago, Chicago, 
Illinois 60637} \\
\r {13}  {\eightit Joint Institute for Nuclear Research, RU-141980 Dubna, Russia}
\\
\r {14} {\eightit Duke University, Durham, North Carolina  27708} \\
\r {15} {\eightit Fermi National Accelerator Laboratory, Batavia, Illinois 
60510} \\
\r {16} {\eightit University of Florida, Gainesville, Florida  32611} \\
\r {17} {\eightit Laboratori Nazionali di Frascati, Istituto Nazionale di Fisica
               Nucleare, I-00044 Frascati, Italy} \\
\r {18} {\eightit University of Geneva, CH-1211 Geneva 4, Switzerland} \\
\r {19} {\eightit Glasgow University, Glasgow G12 8QQ, United Kingdom}\\
\r {20} {\eightit Harvard University, Cambridge, Massachusetts 02138} \\
\r {21} {\eightit The Helsinki Group: Helsinki Institute of Physics; and Division of
High Energy Physics, Department of Physical Sciences, University of Helsinki, FIN-00044, Helsinki, Finland}\\
\r {22} {\eightit Hiroshima University, Higashi-Hiroshima 724, Japan} \\
\r {23} {\eightit University of Illinois, Urbana, Illinois 61801} \\
\r {24} {\eightit The Johns Hopkins University, Baltimore, Maryland 21218} \\
\r {25} {\eightit Institut f\"{u}r Experimentelle Kernphysik, 
Universit\"{a}t Karlsruhe, 76128 Karlsruhe, Germany} \\
\r {26} {\eightit High Energy Accelerator Research Organization (KEK), Tsukuba, 
Ibaraki 305, Japan} \\
\r {27} {\eightit Center for High Energy Physics: Kyungpook National
University, Taegu 702-701; Seoul National University, Seoul 151-742; and
SungKyunKwan University, Suwon 440-746; Korea} \\
\r {28} {\eightit Ernest Orlando Lawrence Berkeley National Laboratory, 
Berkeley, California 94720} \\
\r {29} {\eightit University of Liverpool, Liverpool L69 7ZE, United Kingdom} \\
\r {30} {\eightit University College London, London WC1E 6BT, United Kingdom} \\
\r {31} {\eightit Massachusetts Institute of Technology, Cambridge,
Massachusetts  02139} \\   
\r {32} {\eightit Institute of Particle Physics: McGill University,
Montr\'{e}al, Canada H3A~2T8; and University of Toronto, Toronto, Canada
M5S~1A7} \\
\r {33} {\eightit University of Michigan, Ann Arbor, Michigan 48109} \\
\r {34} {\eightit Michigan State University, East Lansing, Michigan  48824} \\
\r {35} {\eightit Institution for Theoretical and Experimental Physics, ITEP,
Moscow 117259, Russia} \\
\r {36} {\eightit University of New Mexico, Albuquerque, New Mexico 87131} \\
\r {37} {\eightit Northwestern University, Evanston, Illinois  60208} \\
\r {38} {\eightit The Ohio State University, Columbus, Ohio  43210} \\  
\r {39} {\eightit Okayama University, Okayama 700-8530, Japan}\\  
\r {40} {\eightit Osaka City University, Osaka 588, Japan} \\
\r {41} {\eightit University of Oxford, Oxford OX1 3RH, United Kingdom} \\
\r {42} {\eightit University of Padova, Istituto Nazionale di Fisica 
          Nucleare, Sezione di Padova-Trento, I-35131 Padova, Italy} \\
\r {43} {\eightit University of Pennsylvania, Philadelphia, 
        Pennsylvania 19104} \\   
\r {44} {\eightit Istituto Nazionale di Fisica Nucleare, University and Scuola
               Normale Superiore of Pisa, I-56100 Pisa, Italy} \\
\r {45} {\eightit University of Pittsburgh, Pittsburgh, Pennsylvania 15260} \\
\r {46} {\eightit Purdue University, West Lafayette, Indiana 47907} \\
\r {47} {\eightit University of Rochester, Rochester, New York 14627} \\
\r {48} {\eightit The Rockefeller University, New York, New York 10021} \\
\r {49} {\eightit Istituto Nazionale di Fisica Nucleare, Sezione di Roma 1,
University di Roma ``La Sapienza," I-00185 Roma, Italy}\\
\r {50} {\eightit Rutgers University, Piscataway, New Jersey 08855} \\
\r {51} {\eightit Texas A\&M University, College Station, Texas 77843} \\
\r {52} {\eightit Texas Tech University, Lubbock, Texas 79409} \\
\r {53} {\eightit Istituto Nazionale di Fisica Nucleare, University of Trieste/\
Udine, Italy} \\
\r {54} {\eightit University of Tsukuba, Tsukuba, Ibaraki 305, Japan} \\
\r {55} {\eightit Tufts University, Medford, Massachusetts 02155} \\
\r {56} {\eightit Waseda University, Tokyo 169, Japan} \\
\r {57} {\eightit Wayne State University, Detroit, Michigan  48201} \\
\r {58} {\eightit University of Wisconsin, Madison, Wisconsin 53706} \\
\r {59} {\eightit Yale University, New Haven, Connecticut 06520} \\
\end{center}

%
%
\begin{abstract}
The Standard Model predictions for $W\gamma$ and $Z\gamma$ production are 
tested using an integrated luminosity of
$200$ pb$^{-1}$ of \ppbar collision data collected at the Collider 
Detector at Fermilab. The cross sections are measured by selecting 
leptonic decays of the $W$ and $Z$ bosons, and photons with transverse 
energy $E_T>7$ GeV that are well separated from leptons.
The production cross sections and kinematic distributions for the  \Wg and \Zg data are
compared to SM predictions.
\end{abstract}
\pacs{PACS numbers 14.65.Ha, 13.85.Qk, 13.85.Ni}

\twocolumn
The study of the characteristics of \Wg and \Zg production is an important test of the 
Standard Model (SM) description of gauge boson interactions and is sensitive to physics beyond the SM.
The \Wg and \Zg cross sections are directly sensitive to the trilinear gauge couplings 
which are uniquely predicted
by the non-Abelian gauge group of the SM electroweak sector $SU(2)_L\times U(1)_Y$. 
\Wg production can be used to study the $WW\gamma$ vertex and \Zg production can be 
used to constrain the $ZZ\gamma$ and $Z\gamma\gamma$ vertices which vanish in the 
SM~\cite{tgc,prd41_1990_1476,prd47_1993_4889}.
Physics beyond the SM (e.g. compositeness models or excited $W$ or $Z$ bosons)
could alter the cross sections and the production kinematics. 
\Wg and \Zg production are also important background 
contributions to searches for new physics, e.g. in Gauge Mediated Supersymmetry Breaking models~\cite{gmsb}.

This report presents measurements of $p\bar{p} \to l \nu \gamma + X$ and $p\bar{p} \to l^+ l^- \gamma + X$ 
production at $\sqrt{s}$=1.96 TeV at the Tevatron accelerator using data 
obtained with the upgraded Collider Detector at Fermilab (CDF). In the SM the 
$l\nu\gamma$ and $l^+ l^- \gamma$ final states occur due to 
$W\gamma \to l\nu \gamma$ and $Z\gamma \to l^+l^-\gamma$ production, as well as via
lepton bremsstrahlung: $W\to l\nu \to l\nu \gamma$ and $Z \to l^+l^- \to l^+l^- \gamma$. 
Throughout this letter the notation ``$Z$'' is used to specify $Z/\gamma^*$ 
production via the Drell-Yan process. 
The notations \Wg and \Zg are used to denote the $l\nu\gamma$ and $l^+ l^- \gamma$ final states.

The data are taken at higher center of mass energy and constitute a larger data sample 
by at least a factor of two than previous measurements~\cite{d099,d0wg,d0zg,cdfwg,cdfzg}. 
They were collected between March 2002 and September 2003, and correspond to an 
integrated luminosity of about $200$ pb$^{-1}$.
$W$ and $Z$ bosons are selected in their electron and muon decay modes. 
Additionally, a photon with transverse energy above 7 GeV is selected. 
The production properties of the \Wg and \Zg events are compared 
to the SM predictions.


The CDF detector is described in detail elsewhere~\cite{CDF_run2}. 
Transverse momenta of charged particles ($p_T$)\footnote{
We use a cylindrical coordinate system about the beampipe in which $\theta$ is the
polar angle, $\phi$ is the azimuthal angle and $\eta=-\ln \tan(\theta/2)$. $E_T=E\sin\theta$ 
and $p_T = p \sin \theta$ where $E$ is the energy measured by the calorimeter and $p$ the 
momentum measured in the tracking system. $\vec{\met}=-\sum_i E_T^i \vec{n_i}$ 
where $\vec{n_i}$ is a unit vector that points from the interaction vertex to 
the $i$th calorimeter tower in the transverse plane.
$\met$ is the magnitude of $\vec{\met}$. 
If muons are identified in the event, $\met$ is corrected for the muon momenta.}
are measured by an eight-layer silicon strip detector~\cite{SVX} 
and a 96-layer drift chamber (COT)~\cite{COT} inside a 1.4 Tesla magnetic field.
The COT provides coverage with high efficiency for $|\eta|<1$. At higher $|\eta|$
the silicon detector is used for measuring charged particles. Electromagnetic and hadronic calorimeters
surround the tracking system. They are segmented in a projective tower geometry and measure
energies $E$ of charged and neutral particles in the central ($|\eta|<1.1$) and forward ($1.1<|\eta|<3.6$) regions. 
Each calorimeter has an electromagnetic shower profile detector 
positioned at the shower maximum. Located at the inner face of the central calorimeter, 
the central preradiator chambers use the solenoid coil as a radiator 
to measure the shower development. 
These two detectors are used for the photon identification and background
determination.  The calorimeters are surrounded by muon drift chambers 
covering $|\eta|<1$.
Gas Cherenkov counters ~\cite{CLC} measure the average number of $p\bar{p}$ 
inelastic collisions per bunch crossing and thereby determine the beam luminosity.


For the $W$ and $Z$ boson selection
with decays into muons or central electrons, the trigger is solely based on the 
identification of a high transverse momentum lepton~\cite{wzincl}. For $W$'s decaying to forward electrons, 
the trigger additionally requires $\met>$ 15 GeV. Offline, a 
high-\pt~lepton ($l=e,\mu$) is required to fulfill tighter selection 
criteria~\cite{wzincl}. 
Electron candidates are required to have $E_T >$ 25 GeV and $|\eta|<$ 2.6. 
In the central region, a COT track with $p_T >$ 10 GeV/$c$ must be associated with the energy deposition, 
while in the forward region calorimeter-seeded silicon tracking is used to associate a track with the electromagnetic shower~\cite{cigdem}.
The electromagnetic shower profile of an electron candidate must be 
consistent with expectations from test beam data.
Muons are selected by requiring a COT track with $p_T >$ 20 GeV/$c$, 
and the associated energy deposition in the calorimeter to be consistent with
that expected for a muon~\cite{wzincl}.
In addition, for at least one muon per event, the track segments in the muon chambers must match
the extrapolated position of the muon track and be in the range $|\eta| <$ 1.0. 
Both electrons and muons must be isolated from other calorimeter energy depositions~\cite{wzincl}.
The selected samples correspond to 
an integrated luminosity of 202 \invpb~ (168 \invpb~) for central (forward) electrons
and 192 \invpb~ (175 \invpb~) for muons in the region $|\eta|<0.6$ ($0.6<|\eta|<1.0$).

For $W\rightarrow l\nu$ candidates, we also require \met $>$ 25 (20) GeV
in the electron (muon) channel as evidence for the neutrino.
For the $W\rightarrow \mu\nu$ channel, events with an additional track 
with $p_T >$ 10 GeV/$c$ and a calorimeter signal consistent with a muon, are
rejected as potential background from $Z\to\mu^{+}\mu^{-}$.
For the selection of $Z$ candidates, a second electron is required in the 
electron channel and a second isolated track consistent with a minimum 
ionizing particle in the muon channel.

In the \Zg analysis the invariant mass of the dilepton pair, $\mll$, 
is required to be in the range 40 $< \mll <$ 130 GeV/$c^2$ 
to enhance the sensitivity to on-shell $Z$ boson production. 
In the \Wg analysis the transverse mass, $M_T(l,\met)$,
is required to be in the range 30 $< M_T(l,\met) <$ 120 GeV/$c^2$ to select
on-shell $W$ boson production. 
The transverse mass is used since the longitudinal component 
of the neutrino momentum cannot be measured: 
$M_T(l,\met)=\sqrt{2 p_T^l \mbox{${\hbox{$E$\kern-0.6em\lower-.1ex\hbox{/}}}_T$} 
(1-\cos\phi_{l,\met})}$, 
where $\phi_{l,\met}$ is the difference in azimuthal angle between the lepton momentum 
and the missing transverse momentum vector. 

After reconstructing a $W$ or $Z$ candidate, we select a 
photon with $E_T^\gamma >$ 7 GeV within $|\eta_\gamma| <$ 1.0 which is
isolated from other particles in both the calorimeter and the tracking 
detectors. 
The transverse energy deposit around the photon in a cone 
$\Delta R=\sqrt{(\eta_i-\eta_\gamma)^2 + (\phi_i-\phi_\gamma)^2} =$ 0.4
is required to be less than $10\%$ of the photon transverse 
energy for $E_T^\gamma <$ 20 GeV
and less than $2+0.02(E_T^\gamma-20)$ GeV for $E_T^\gamma >$ 20 GeV. Here, $\eta_i$ and $\phi_i$ denote
the location of the energy deposit in the $i$th calorimeter tower excluding those associated
with the photon candidate. The total sum of the track transverse
momenta in a cone of $0.4$ around the photon candidate is also 
required to be less than $2$ GeV/$c$.
To remove electron background we require there to be no track with $p_T >$ 1 GeV/$c$
pointing toward the photon candidate.
The photon candidate also must have a shower shape consistent with a 
single particle and must be separated from the lepton by
$\Delta R(l,\gamma)=\sqrt{(\eta_l-\eta_\gamma)^2 + (\phi_l-\phi_\gamma)^2} >$ 0.7.
This last requirement is placed to suppress the contribution from 
bremsstrahlung photons. 
After all selection criteria are applied, 323 $W\gamma$ candidates
and 71 $Z\gamma$ candidates are found.

The most important source of background to both the $Z\gamma$ and $W\gamma$
analysis is the production of a real $Z$ or $W$ boson and a hadron which is 
misidentified as a photon. 
This background is determined using
large event samples triggered on jets at several $E_T$ thresholds: 20, 50, 
70 and 100 GeV. We measure the fraction of jets in the samples which pass
all the photon selection requirements. This fraction is then corrected
for prompt photon contamination within the jet samples. Two methods were used
to estimate this contamination. For $E_T^\gamma<40$ GeV, the estimate
of the prompt photon contamination exploits the broader shower 
shape of $\pi^0 \to \gamma \gamma$ showers compared to prompt $\gamma$ 
showers in the electromagnetic shower profile detector \cite{promptgammaprd}. 
For $E_T^\gamma>40$ GeV hits in the 
central preradiator chambers are counted. In this method prompt photons are distinguished 
from meson decays since
the probability of a photon conversion in the magnetic coil is higher for $\pi^0$'s than for prompt
photons \cite{promptgammaprd}. The resulting fake rate for a jet to pass all photon selection cuts
is about 0.3\% at $E_T=10$ GeV and decreases exponentially to about 0.07\% for $E_T=25$ GeV. 

We obtain the background prediction by applying this fake rate to jets in $W$ and $Z$ events.  
The background due to events where neither the leptons nor the photon are genuine 
is implicitly taken into account in the above estimate. 
In the $W\gamma$ analysis an additional background arises 
from $Z\gamma$ production
where large $\met$ is observed due to an undetected lepton.
This background is larger in the muon than in the electron
channel due to the smaller muon coverage of the CDF detector. Another source of background is
$\tau \nu_\tau \gamma \to l \nu_l \bar{\nu_\tau} \nu_\tau \gamma$ production.
These two backgrounds are determined using the Monte Carlo generators described below.
$\tau \tau \gamma$ is found to be a negligible source of background in both analyses.

A summary of the background contributions for the $W\gamma$ analysis is given in Table \ref{tab:wbg}.
For the $Z\gamma$ analysis, the only background is due to jets mis-identified as photons.  
For $ee\gamma$, the estimated background is $2.8 \pm 0.9$ events,
and for $\mu\mu\gamma$, it is $2.1 \pm 0.6$ events.
\begin{table}[htbp]
\begin{center}
\caption{\label{tab:wbg} Background event contributions 
for the $e \nu \gamma$ and 
$\mu \nu \gamma$ analyses. The combined statistical and systematic 
uncertainty on the background prediction is also quoted.}
\begin{tabular}{|l|r|r|}\hline
                              & $e \nu \gamma$ & $\mu \nu \gamma$\\ \hline
$W$+jet                       & $59.5 \pm 18.1$ & $27.6 \pm 7.5$  \\
$\tau \nu \gamma$ & $ 1.5 \pm  0.2$ & $ 2.3 \pm 0.2$  \\
$l^+ l^- \gamma$              & $ 6.3 \pm  0.3$ & $17.4 \pm 1.0$ \\\hline
Total Background              & $67.3 \pm 18.1$ & $47.3 \pm 7.6$ \\\hline
\end{tabular}
\end{center}
\end{table}

The $p\bar{p}\to l\nu\gamma X$ and $p\bar{p}\to l^+l^-\gamma X$ 
SM signal predictions are determined using leading order
Monte Carlo generators for all three lepton generations. 
The matrix element generator ~\cite{prd41_1990_1476,prd47_1993_4889} includes
initial and final state photon radiation and the $WW\gamma$ vertex diagram. 
Initial state QCD radiation and hadronization are included using {\tt PYTHIA}~\cite{pythia}. 
The parton momentum distribution is modeled with CTEQ5L parton density functions (PDF's)~\cite{cteq5l}.
${\cal O}(\alpha_s)$ QCD corrections~\cite{Baur:1993ir} to the $W\gamma$ and $Z\gamma$ 
production cross sections are calculated using CTEQ5M PDF's~\cite{cteq5l}.
These corrections increase the $W\gamma$ ($Z\gamma$) cross section 
by $33-55\%$ ($27-32\%$) for $E_T^\gamma$ in the range $10-55$ GeV.

The SM cross section for $p\bar{p}\to l\nu\gamma X$ 
production for the kinematic region $E_T^\gamma > 7$ GeV and 
$\Delta R(l,\gamma)>0.7$ is $19.3 \pm 1.4$ pb for the $W$-boson decaying into a single lepton flavor.
For the same kinematic region and with the invariant mass of the 
dilepton pair $\mll>$ 40 GeV/$c^2$, 
the cross section for $p\bar{p}\to l^+l^-\gamma X$ 
production is $4.5 \pm 0.3$ pb for the $Z$-boson decaying
into a lepton pair of a single flavor. 
The 7\% uncertainty on the cross section 
due to higher order contributions and uncertainties on the PDF's
is evaluated by changing the factorization scale (2\%), the renormalization scale (3\%) 
and comparing the predictions made with several PDF's (5\%)~\cite{cteq6,mrst2002}.

The observed ($N_{obs}$) and expected numbers of signal ($N_{sig}$) and
background ($N_{bg}$) events
in the \Wg and \Zg analyses are given in Table~\ref{tab:wgamma}. 
Both the electron and muon data are in good agreement with expectations.
The systematic uncertainties on these measurements include uncertainties
on the event selection efficiency and acceptance. The main contributions
come from higher order QCD corrections to the acceptance and the efficiency of the photon selection.
The dominant uncertainty on the background is due to the jet fake rate uncertainty.
The total systematic uncertainty on the cross sections is
9-14\% for the \Wg and 3\% for the \Zg cross sections. An additional
uncertainty of 6\% arises from the luminosity measurement.
\begin{table}[htbp]
\begin{center}
\caption{\label{tab:wgamma} Expected and observed numbers of events for
$e \nu \gamma$ and $\mu \nu \gamma$, $e^+e^-\gamma$ and $\mu^+\mu^-\gamma$ production.
The systematic uncertainties listed for the expected number of events 
excludes the 7\% uncertainty on the theoretical 
cross section and the 6\% uncertainty in luminosity measurement.
The product of the acceptance and efficiency, $A\times \epsilon$, and the 
measured cross sections, $\sigma(l\nu\gamma)$ and $\sigma(l^+l^-\gamma)$, are also listed. The 
first uncertainty is statistical and the second is systematic. 
There is a separate error
on the luminosity normalization of $1.2$ pb for the \Wg and $0.3$ pb for the \Zg cross
section measurements.}
\begin{tabular}{|l|l|l|}\hline
                      & $e\nu\gamma$ & $\mu\nu\gamma$ \\\hline
$N_{sig}$             & $126.8 \pm 5.8 $  & $95.2 \pm 4.9$ \\
$N_{sig}+N_{bg}$      & $194.1 \pm 19.1 $ & $142.4 \pm 9.5$  \\
$N_{obs}$             & $195$  & $128$\\\hline
$A\times \epsilon$    & $3.3\%$  & $2.4\%$\\\hline
$\sigma(l \nu \gamma)$ (pb)  & $19.4 \pm 2.1 \pm 2.9$ & $16.3\pm 2.3 \pm 1.8$ \\\hline
\hline
              & $e^+ e^- \gamma$ & $\mu^+ \mu^- \gamma$ \\\hline
$N_{sig}$     & $31.3 \pm 1.6 $& $33.6 \pm 1.5 $ \\
$N_{sig}+N_{bg}$ & $ 34.1 \pm   1.8 $& $ 35.7 \pm 1.7 $ \\
$N_{obs}$     & $36$             & $35$                \\\hline
$A\times \epsilon$& $3.4\%$  & $3.7\%$\\\hline
$\sigma(l^+ l^- \gamma)$ (pb) & $4.8 \pm 0.8 \pm 0.3$ & $4.4 \pm 0.8 \pm 0.2$ \\\hline
\end{tabular}
\end{center}
\end{table}

The cross section $\sigma(l \nu \gamma)$ is 
measured in the kinematic range $\Delta R(l,\gamma)>0.7$ 
and $E_T^\gamma >7$ GeV 
with $\sigma = (N_{obs}-N_{bg})/(A \times \epsilon \times \intlumi)
=(N_{obs}-N_{bg})/N_{sig} \times \sigma_{SM}$.
Here, $\intlumi$ is the integrated luminosity, 
$A$ is the acceptance, $\epsilon$
is the selection efficiency and $\sigma_{SM}$ is the SM cross section of the
Monte Carlo simulation sample used for estimating 
the acceptance and number of expected signal events.
The resulting cross sections are given in Table~\ref{tab:wgamma}.
The measured cross sections are determined for the full
$W$ decay phase space, transverse mass range and photon $\eta$ 
range using extrapolations based upon the SM 
expectation~\cite{Baur:1993ir}.
Combining the electron and muon channel, 
assuming lepton universality, and
taking into account correlations of the systematic uncertainties, yields 
$\sigma(l \nu \gamma )=18.1 \pm 3.1$ pb.
The theoretical prediction for this cross section is
$19.3 \pm 1.4$ pb.

The cross section $\sigma(l^+ l^- \gamma)$ is measured
in the kinematic range  $\Delta R(l,\gamma) >$ 0.7, $E_T^\gamma >$ 7 GeV and $\mll >$ 40 GeV/$c^2$. 
We follow the same procedure as for the \Wg analysis 
and obtain the cross section listed in Table~\ref{tab:wgamma}.
The measured cross sections are determined for the full
$Z$ decay phase space, dilepton mass range $\mll>$ 40 GeV/$c^2$ and photon $\eta$ range using extrapolations
based upon the SM expectation~\cite{Baur:1993ir}. 
The combined electron and muon result is $\sigma(l^+ l^- \gamma)=4.6 \pm 0.6$ pb.
The theoretical prediction for this cross section is
$4.5 \pm 0.3$ pb.
\begin{figure}[htbp]
\includegraphics[width=7.5cm]{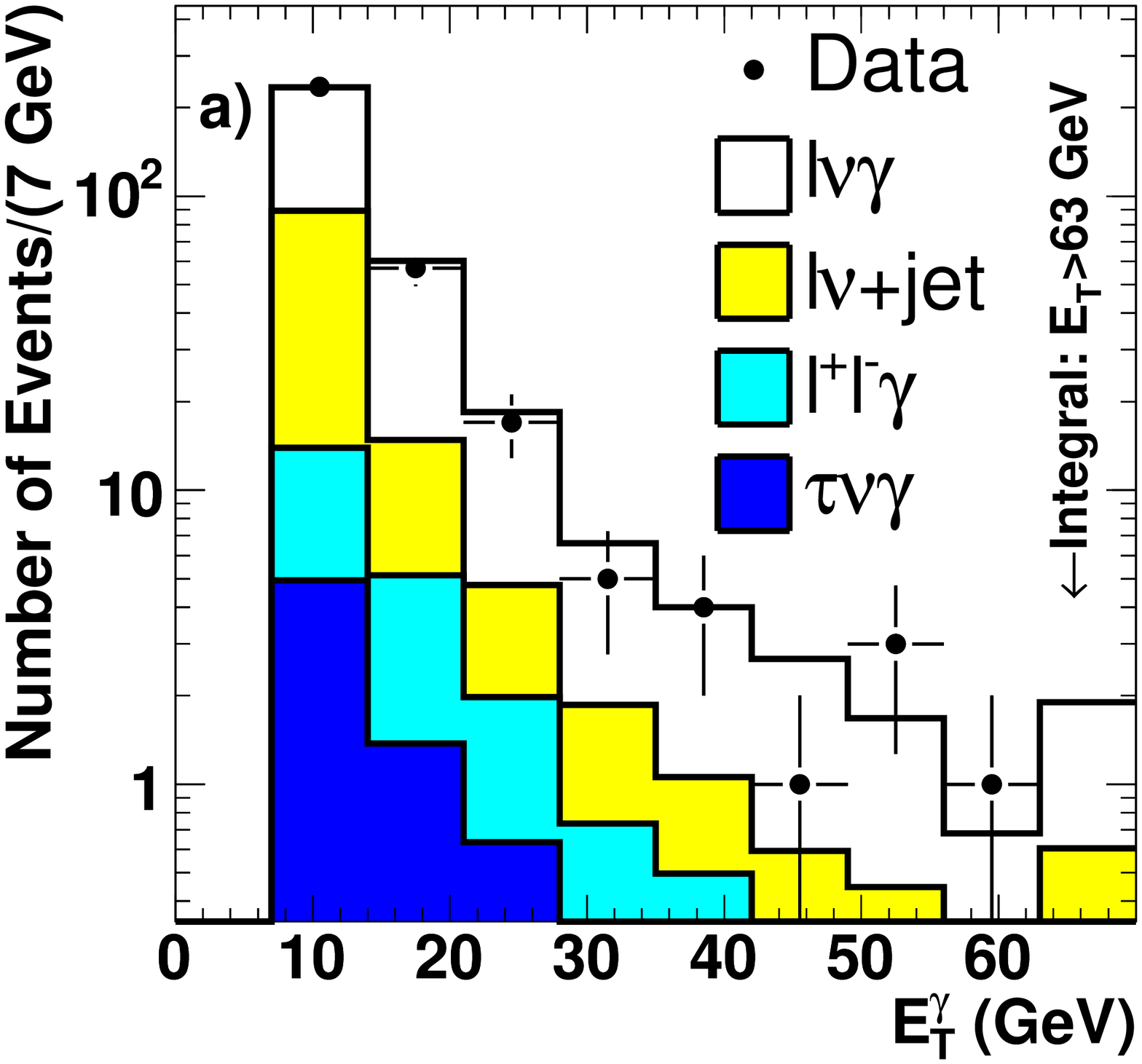}
\includegraphics[width=7.5cm]{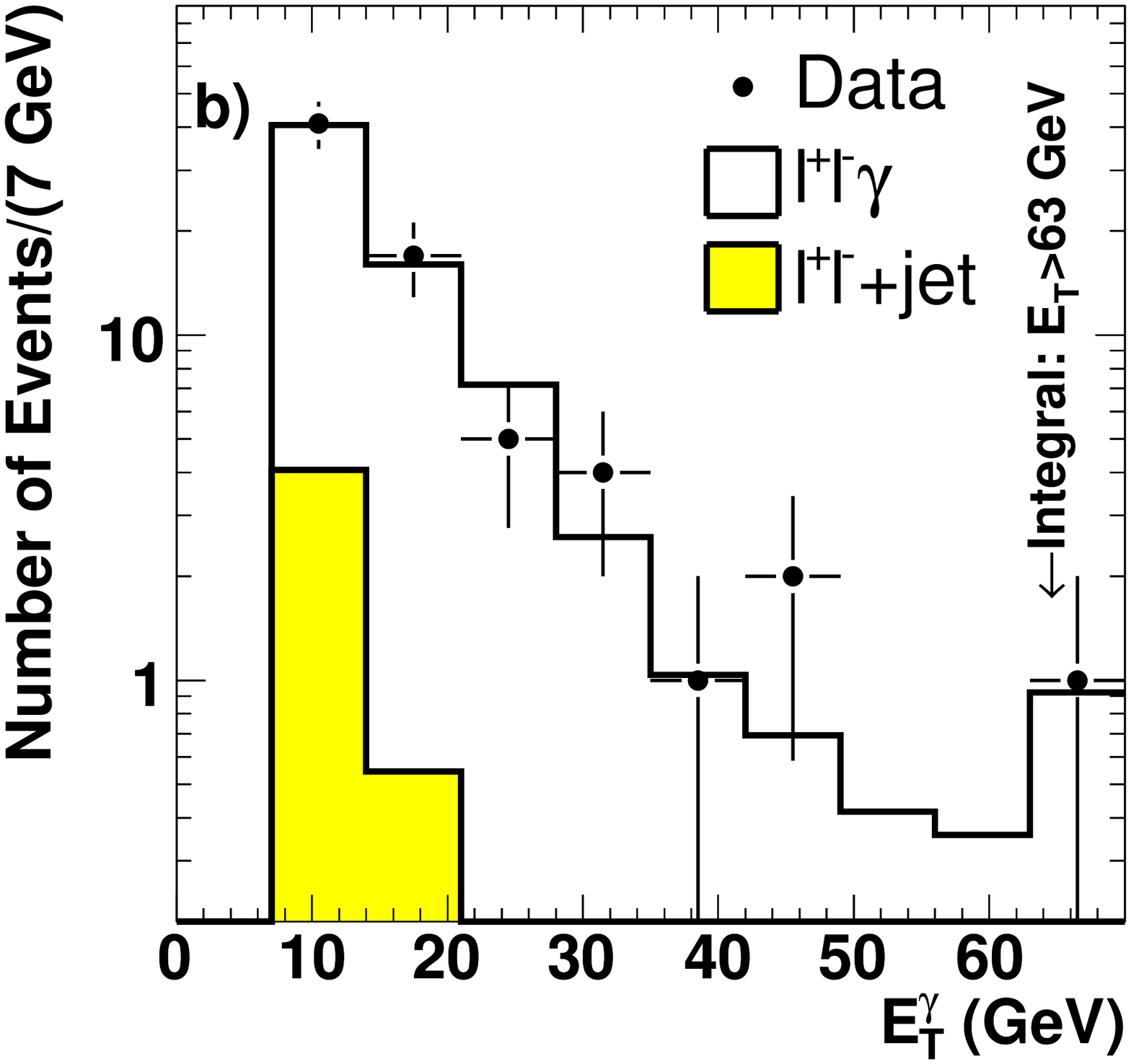}
\caption{Photon transverse energy spectrum, $E_T^\gamma$, 
for a) \Wg and b) \Zg candidates selected in the leptonic decay channel.  
The data are compared with the SM expectations for signal and background with
the histograms added cumulatively. In both figures 
the last bin contains all events with $E_T^\gamma >$ 63 GeV.} 
\label{fig:wgamet}
\end{figure}

\begin{figure}[htbp]
\includegraphics[width=7.5cm]{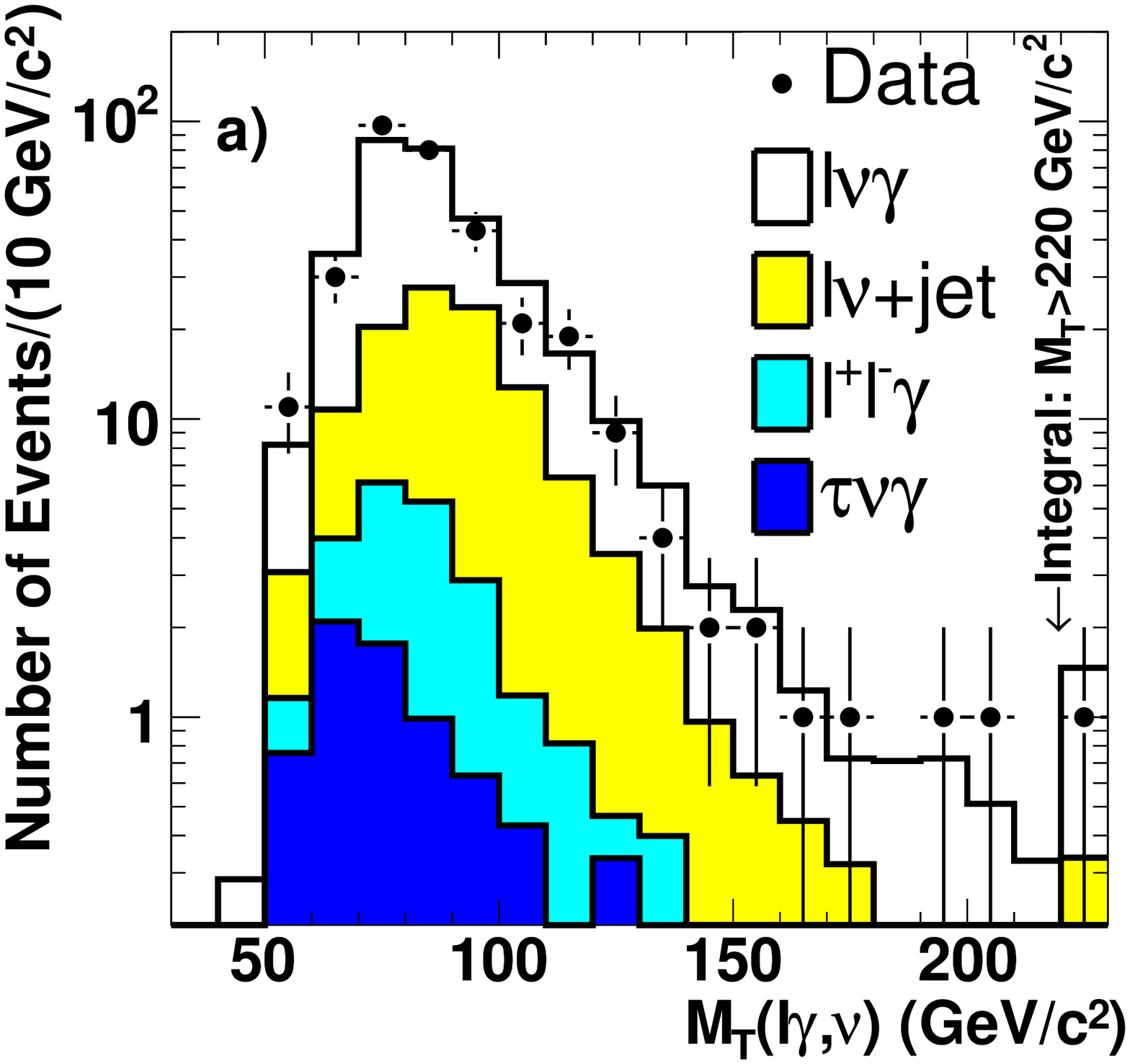}
\includegraphics[width=7.5cm]{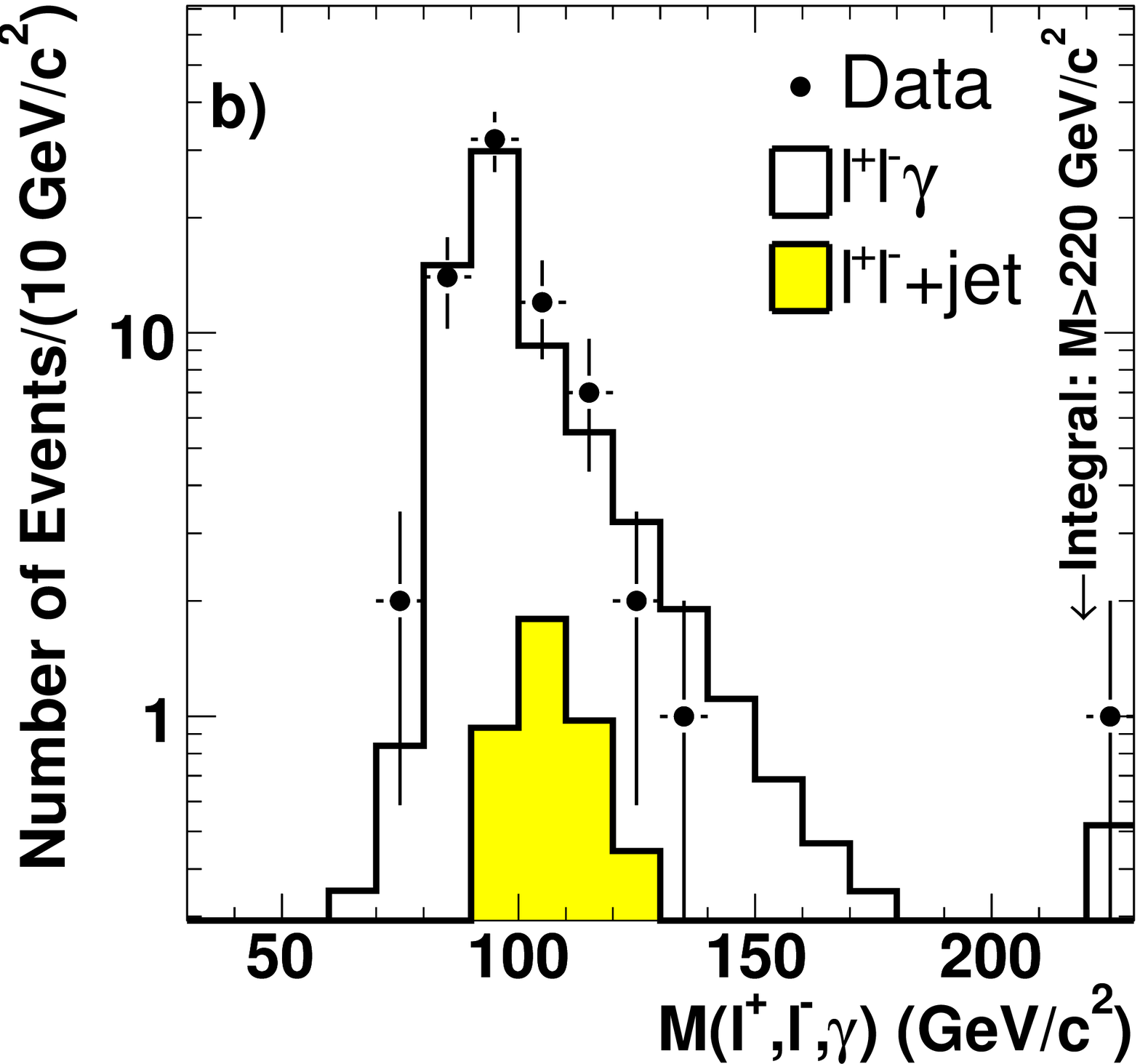}
\caption{a) The cluster transverse mass of the lepton-photon-missing 
$E_T$ system for \Wg candidates, and b) the invariant mass of 
the lepton-lepton-photon system for \Zg candidates.  
The data are compared with the SM expectations for signal and background
with the histograms added cumulatively. In both figures the last bin contains all events 
with masses above $220$ GeV$/c^2$.}
\label{fig:ctm}
\end{figure}

In addition to these cross section measurements,
we compare the SM predictions for several kinematic variables with the data 
for $E_T^\gamma>7$ GeV and $\Delta R(l,\gamma)>0.7$. 
We choose the  $E_T^\gamma$ and final state mass spectra since these are
sensitive tests of SM predictions.
The transverse energy of the photon in \Wg and \Zg production is shown in 
Figure~\ref{fig:wgamet}. 
Figure~\ref{fig:ctm} shows the cluster transverse mass~\cite{CTM}, 
$M_T(l\gamma,\nu)$, for \Wg events and the invariant mass of the 
($l^+,l^-,\gamma$) system, $M(l^+,l^-,\gamma)$, for \Zg events. 
The data are in good agreement with the SM expectations for both
processes. The event with the highest $E_T$ photon, observed in the $e^+e^-\gamma$ channel with $E_T^\gamma$ = 141 GeV and $M(e^+,e^-,\gamma)$ = 382 GeV$/c^2$, is 
consistent with the rate expected from SM predictions.




In summary, we have measured \Wg and \Zg production 
in $p\bar{p}$ collisions at $\sqrt{s}$ = 1.96 TeV using 
data from the CDF experiment. 
The cross sections, measured
to a precision of 15\%, are compared to electroweak predictions
having an estimated uncertainty of 7\%. 
For $E_T^\gamma$ above 7 GeV and $\Delta R(l,\gamma)>0.7$, the
production cross sections, and the photon and $W/Z$ boson production kinematics,
are found to agree with SM predictions. 

\begin{center}
\textbf{Acknowledgments}
\end{center}

We thank the Fermilab staff and the technical staffs of the participating
institutions for their vital contributions.  We also thank Ulrich Baur for his 
help and input toward the understanding of theoretical aspects of this analysis.

This work was supported by the
U.S. Department of Energy and National Science Foundation; the Italian
Istituto Nazionale di Fisica Nucleare; the Ministry of Education, Culture,
Sports, Science and Technology of Japan; the Natural Sciences and
Engineering Research Council of Canada; the National Science Council of the
Republic of China; the Swiss National Science Foundation; the A.P. Sloan
Foundation; the Bundesministerium f\"{u}r Bildung und Forschung, Germany; the
Korean Science and Engineering Foundation and the Korean Research
Foundation; the Particle Physics and Astronomy Research Council and the
Royal Society, UK; the Russian Foundation for Basic Research; the Comisi\'on
Interministerial de Ciencia y Tecnolog\'{\i}a, Spain; and in part by the European
Community's Human Potential Programme under contract HPRN-CT-20002, Probe
for New Physics.

\end{document}